\documentclass[namedreferences]{SolarPhysics}

\usepackage[optionalrh]{spr-sola-addons} 
\usepackage{graphicx}        
\usepackage{color}           
\usepackage{url}             
\usepackage{rotating}

\newcommand{\etal}{{\it et al.}}

\begin{document}

\begin{article}

\begin{opening}

\title{A Statistical Survey of Hard X-ray Spectral Characteristics of Solar Flares with Two Footpoints.}

\author{P.~\surname{Saint-Hilaire}$^{1}$\sep
        S.~\surname{Krucker}$^{1}$\sep
        R.P.~\surname{Lin}$^{1}$      
       }
\runningauthor{Saint-Hilaire \etal}
\runningtitle{Hard X-ray Double Footpoints: a Statistical Survey}

   \institute{$^{1}$ Space Sciences Laboratory, UC Berkeley
                     email: \url{shilaire@ssl.berkeley.edu} email: \url{krucker@ssl.berkeley.edu}\\ 
             }
\begin{abstract}

	Using RHESSI data, we have analyzed some 172 hard X-ray peaks during 53 solar flares which exhibited a double-footpoint structure.
	Fitting both footpoints with power-laws, we find that spectral index differences $\Delta\gamma$ range mostly between 0 to 0.6, and only rarely go beyond.
	Asymmetries between footpoints were not observed to be significantly dependent on their mean heliographic position, their relative position with respect to each other,
	nor their orientation with respect to the solar equator.	
	Assuming a symmetric acceleration process, it is also clear that differences in footpoint spectral indices and footpoint flux ratios can seldom be attributed to a difference in column densities between the two legs of a coronal loop. 
	Our results corroborate better the magnetic mirror trap scenario.
	Moreover, footpoint asymmetries are more marked during times of peak HXR flux than when averaging over the whole HXR burst, 
	suggesting that the magnetic configuration evolves during individual HXR bursts.
	We observed also a linear correlation between the peak 50-keV flux and the peak GOES 1--8{\AA} channel flux, 
	and that HXR burst duration seem correlated with loop length.

\end{abstract}


\end{opening}
\section{Introduction}\label{sect:intro}

Solar flares are believed to be powered by magnetic reconnection high in the corona, which accelerate particles.
Particles, and in particular electrons, travel down field lines and emit bremsstrahlung hard X-rays (HXR) as they penetrate the denser chromosphere.
Hence, the HXR radiation from the footpoints contain much information about the accelerated electrons (though convoluted with transport effects), such as energy content 
\cite{Brown1971,Holman2003,PSH2005}.

The first observations of HXR footpoints were made by \inlinecite{Hoyng1981} on SMM.
Yohkoh HXT later characterized the ``standard'' flare model: two footpoints and an above-the-looptop source \cite{Masuda1994}.
The above-the-looptop source is rarely observed, although that could be due to observational constraints.

Statistical studies of HXR footpoints (and looptop) sources' spectral indices using the Yohkoh satellite \cite{Sakao1994,Petrosian2002} have been carried out in the past.
Some of their results seem to indicate that the spectral indices of two neighbouring footpoints could differ by as much as 1 or even 2!
Such differences cannot be explained easily by transport mechanisms (see {\it e.g.} Appendix \ref{appendix:ColumnDensity}).
Yohkoh HXT's results were compromised by the fact that it had only had 4 energy channels, and had to deal with issues like thermal contamination of the lowest channel(s), and the sometime poor statistics of the upper ones.

The Ramaty High Energy Solar Spectroscopic Imager (RHESSI, \opencite{Lin2002}) offers an unprecedented combination of spectral resolution (1 keV in the 3--100 keV range), spatial resolution (2.3$''$), temporal evolution ($\approx$2s), and sensitivity comparable to Yohkoh HXT's.
Previous RHESSI imaging spectroscopy results include: \inlinecite{Krucker2002}, \inlinecite{Emslie2003}, \inlinecite{Marina2006}.
A similar time variation of spectra in footpoints is observed, although in some cases, the spectrum in one footpoint is steeper than in the other one (by about 0.3 in spectral index, in \inlinecite{Emslie2003}.
\inlinecite{Emslie2003} suggest that the discrepancy could be due to a difference in column densities of the electron population, as they propagate down an asymmetric loop (cf Appendix \ref{appendix:ColumnDensity}).
	
All RHESSI papers so far discuss only a single event, or a few events, but no statistical study of the footpoints has been done so far, that exploit RHESSI's large database of observed flares.
These results should help constrain energy release and particle acceleration in the flare model.

\section{Event Selection}\label{sect:selection}

			\begin{figure}
			\centering
			\includegraphics[width=11cm]{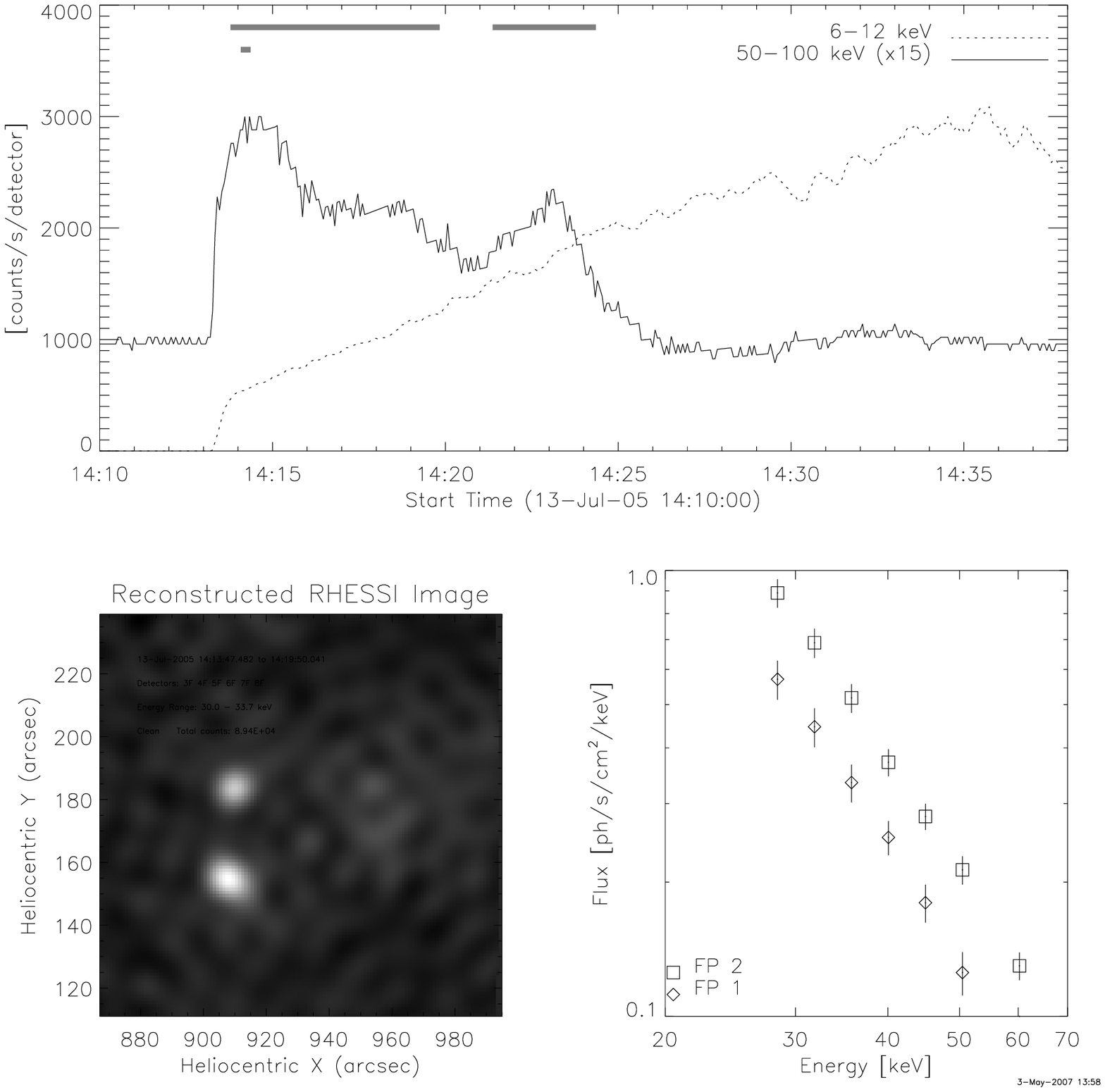}
			\caption{RHESSI time profile (top), 30--33 keV image (bottom left) and footpoint spectra (Bottom right) of the 13 July 2005 M2.7 flare.
				The gray bars represent the time intervals chosen for our analysis: a ``Peak Times'' 16-second interval around 14:14:10, and two ``Whole Peaks'' 
				from 14:13:47 to 14:19:50, and from 14:21:22 to 14:24:41. 
				The third HXR peak ($\approx$14:30) was discarded as it was faint, and there was an attenuator state change (Lin \etal, 2002) during it, further complicating the analysis. 
				The sun was eclipsed by the Earth until $\approx$14:13.}
			\label{fig:exampleflare}
			\end{figure}

RHESSI has been in orbit for six years now. 
Since its launch on 5 February 2002, it has observed more than 20000 flares.
For this study, the strongest flares will be taken into considerations, as they have better count statistics.
More specifically, flares which show substantial emission above 50 keV (where solar flare thermal components are always negligible), enough to produce images of good enough quality to be used in imaging spectroscopy (2000 counts is about the minimum for a reliable single image reconstruction with two sources).
To achieve our objectives, we will use the simplest possible events, those that show only two footpoints.

	The HESSI Experimental Data Center (HEDC, \opencite{PSH2002}) was used to find our events.
	It was queried for all flares between 13 February 2002 and 1 July 2006 which had 2 or more sources, and with peak GOES flux above M1.0 level.
	More than 1100 flares corresponded to that description.
	Each were individually examined, in particular the time vs. energy panels of RHESSI images that HEDC automatically produces for each flare (one minute accumulations over whole flare duration).
	The flares that were retained were those that visually displayed two footpoints in HEDC images above 50 keV.
	53 flares were kept (see Table \ref{table:1}).
	
 	\begin{table}
	\centering
	\begin{tiny}
	\caption{Flares studied, GOES X-ray class, and position on the sun in both arcseconds from suncenter and heliographic coordinates.}
	\label{table:1}
	\begin{tabular}{c c c c c c} 
		\hline
		Flare & Date and& GOES	& \multicolumn{3}{c}{Location on Sun}  \\
		number&	time	& class	& X [$''$] & Y[$''$] & Lat/Lon \\
		\hline
1 & 2002/02/20 11:06:12 & C7.5 & 904.4 & 261.5 & N13 W73 \\
2 & 2002/03/17 10:15:36 & M1.4 & -342.2 & -239.2 & S20 E22 \\
3 & 2002/03/17 19:28:44 & M4.4 & -264.8 & -232.9 & S20 E17 \\
4 & 2002/04/10 12:28:04 & M8.8 & -20.8 & 434.5 & N20 E01 \\
5 & 2002/04/10 19:02:48 & M1.8 & -348.2 & 377.5 & N17 E22 \\
6 & 2002/05/31 00:07:08 & M2.4 & -817.5 & -475.4 & S30 E87 \\
7 & 2002/06/01 03:53:40 & M1.6 & -414.8 & -293.0 & S18 E27 \\
8 & 2002/06/02 11:44:32 & M1.0 & -148.8 & -300.3 & S18 E09 \\
9 & 2002/07/17 07:02:48 & M9.2 & 288.5 & 246.5 & N19 W18 \\
10 & 2002/07/18 03:32:56 & M2.5 & 421.4 & 264.5 & N20 W28 \\
11 & 2002/07/23 00:28:04 & X5.1 & -868.1 & -235.3 & S12 E70 \\
12 & 2002/07/29 10:39:08 & M5.1 & 238.9 & -291.3 & S12 W14 \\
13 & 2002/07/31 01:48:40 & M1.4 & 558.4 & -220.2 & S08 W36 \\
14 & 2002/08/03 19:04:36 & X1.2 & 899.8 & -265.8 & S15 W80 \\
15 & 2002/08/21 01:39:16 & M1.6 & 689.2 & -246.7 & S10 W47 \\
16 & 2002/08/22 01:52:00 & M5.9 & 798.4 & -266.5 & S12 W59 \\
17 & 2002/09/08 01:39:08 & M1.6 & -908.8 & -193.1 & S09 E75 \\
18 & 2002/09/27 03:34:28 & M1.0 & -694.7 & 142.9 & N13 E47 \\
19 & 2002/12/04 22:47:00 & M2.5 & -836.7 & 227.7 & N13 E61 \\
20 & 2003/04/23 01:01:56 & M5.1 & 261.8 & 366.8 & N17 W16 \\
21 & 2003/05/29 01:04:40 & X1.2 & 494.4 & -106.6 & S07 W31 \\
22 & 2003/06/17 22:53:40 & M6.8 & -790.4 & -138.6 & S07 E57 \\     
23 & 2003/07/17 08:19:46 & C9.7 & -206.9 & 178.2 & N15 E13 \\
24 & 2003/10/23 08:47:20 & X5.4 & -904.8 & -317.4 & S18 E81 \\
25 & 2003/10/24 02:48:32 & M7.7 & -865.1 & -341.1 & S19 E71 \\
26 & 2003/10/29 20:43:20 & X10 & 90.8 & -381.3 & S18 W05 \\
27 & 2003/11/01 22:33:04 & M3.3 & 818.6 & -253.8 & S13 W60 \\
28 & 2003/11/03 09:49:16 & X3.9 & 917.8 & 130.3 & N08 W73 \\
29 & 2003/11/04 19:33:56 & M9.1 & 900.8 & -335.0 & S19 W81 \\
30 & 2004/01/06 06:22:32 & M2.7 & -972.6 & 88.3 & N05 E89 \\
31 & 2004/01/07 10:22:12 & M3.7 & -930.2 & 117.0 & N05 E73 \\
32 & 2004/04/06 13:22:48 & M2.3 & -261.5 & -170.6 & S16 E16 \\
33 & 2004/07/13 00:15:26 & M6.8 & 654.7 & 181.6 & N14 W45 \\
34 & 2004/07/23 21:19:28 & M1.8 & 125.5 & 6.6 & N05 W07 \\
35 & 2004/09/12 00:33:44 & M4.8 & -706.2 & -35.9 & N02 E47 \\
36 & 2004/10/30 03:30:09 & M3.5 & 316.8 & 145.2 & N12 W19 \\
37 & 2004/10/30 16:24:26 & M6.0 & 427.6 & 139.5 & N12 W26 \\
38 & 2004/10/31 05:32:03 & M2.4 & 540.6 & 152.6 & N12 W34 \\
39 & 2004/11/03 03:30:52 & M1.6 & -674.1 & 94.7 & N08 E44 \\
40 & 2004/11/06 00:30:48 & M9.5 & -79.3 & 83.4 & N08 E04 \\
41 & 2004/11/06 01:42:34 & M3.7 & -27.4 & 68.1 & N07 E01 \\
42 & 2004/11/10 02:09:44 & X2.6 & 700.1 & 91.4 & N07 W46 \\
43 & 2004/12/01 07:10:16 & M1.2 & -335.8 & 128.5 & N08 E20 \\	   
44 & 2005/01/15 06:28:31 & M8.4 & -106.1 & 295.5 & N12 E06 \\
45 & 2005/01/15 22:48:24 & X2.7 & 103.1 & 306.7 & N13 W06 \\
46 & 2005/01/17 10:00:23 & X3.9 & 430.0 & 292.3 & N13 W26 \\
47 & 2005/01/19 08:12:40 & M8.7 & 708.9 & 283.6 & N13 W48 \\
48 & 2005/01/19 10:21:08 & M2.5 & 679.6 & 339.9 & N16 W46 \\
49 & 2005/01/20 06:44:44 & X7.1 & 818.5 & 256.0 & N12 W59 \\
50 & 2005/07/13 14:14:14 & M2.7 & 909.4 & 168.9 & N11 W78 \\
51 & 2005/08/22 01:11:54 & M2.3 & 717.8 & -248.7 & S10 W50 \\
52 & 2005/08/22 17:07:34 & M5.2 & 801.4 & -241.6 & S11 W59 \\
53 & 2005/08/23 14:46:21 & M2.7 & 883.8 & -219.0 & S11 W71 \\
	\hline
	\end{tabular}
	\end{tiny}
	\end{table}

\section{Method of Analysis}\label{sect:method}
	Imaging spectroscopy using CLEAN \cite{Hurford2002} and the OSPEX spectral analysis software were employed.
	Imaging was done using collimators 3 to 8, yielding a formal image resolution of 7$''$ FWHM.
	The time intervals and energy intervals were chosen as follow.

	For each flare, two types of time intervals were used:
	\begin{itemize}
		\item	{\it ``Peak flux'' time intervals:}
			These are four RHESSI spin periods long (each RHESSI spin period being about 4s long), centered at the time of peak HXR (above 50 keV) flux.
			The later is found using RHESSI Observing Summary data \cite{Schwartz2002}.
			The peak flux time interval was taken to be this time of peak flux plus or minus two RHESSI spin periods (which are $\approx$4s long).
			Of course, there can be only one such peak flux time interval per flare, resulting in 53 such peaks in our study.
			Taking the flare of 13 July 2005 as an example (Figure \ref{fig:exampleflare}), the time interval of accumulation was about 14:14:05 to 14:14:22.
		\item	{\it ``Whole peak'' time intervals:}
			Strong non-thermal peaks appearing in RHESSI spectrograms (or dynamic spectra) were selected over their whole {\it time interval}, 
			defined as the time the HXR flux ($>$50 keV) is greater than 50\% of its peak value.
			There can be many such ``Whole peak'' time intervals within the same flare.
			And of course, for each flare, one of the ``Whole peak'' time interval envelops that flare's ``Peak flux'' time interval.
			Again taking the flare of 13 July 2005 as an example (Figure \ref{fig:exampleflare}), the time intervals of accumulation were about 
			14:13:47 to 14:19:50, and 14:21:22 to 14:24:41.
			In a few cases, time intervals did not contain {\bf two} footpoints (but only one, or sometimes three or more), and were hence discarded from the study.
	\end{itemize}

	For each time interval, the energy binning was chosen using the following semi-empirical approach:
	\begin{itemize}
		\item	The {\it start (lowest)} energy was visually chosen by inspection of the RHESSI spectrogram: it is taken to be the point where the non-thermal emission starts to be clearly stronger than the thermal component.
		\item	The energy binning was taken to be pseudo-logarithmic, which each energy bin having at least 2000 counts above background, and the bin width being between 5\% and 20\% of the bin value.
			This was crudely approximated using Observing Summary 4 second data rates.
			The {\it end (highest)} energy was taken to be when the next energy bin could not achieve 2000 counts above background.
			We had 4 to 18 (typically 10) energy bins to fit and obtain spectral indices and fluxes with.
	\end{itemize}

	Finally, for the results presented and discussed in this paper, only fittings deemed ``most reliable'' were kept.
	``Most reliable'' meaning those which had at least 6 or more energy bins for footpoint spectral fitting, and a with a best-fit $\chi^2$ value of 5 or less.
	Ultimately, 33 ``Peak Time'' and 89 ``Whole Peaks'' events were used to produce the results that we analyse here.

	To limit {\it pulse pile-up} issues \cite{Smith2002}, care has been taken to discard times with high count rates ({\it i.e.} just before shutters moving in).
	Moreover, as spectral fitting was usually done above 25 keV, only times with very strong emission (during which both attenuators are ``in'', or ``A3'' state) can potentially produce an additional component around 35 keV
	(with only the thin shutter in (``A1'' state), detector countrates peak around 12 keV, and these can be pile-up to $\approx$24 keV.
	In A3 state, the peak of the response is around $\approx$18 keV counts. These photons can pile up and appear as $\approx$36 keV photons).
	In 20 of our events were the contribution of pile-up photons in certain energy channels larger than 15\%.
	As pile-up typically makes two thermal photons appear as a single higher energy photon, imaging piled-up photons would place them at the location of the thermal source.
	In 19 of these 20 cases, the thermal source was spatially distinct from the HXR footpoints, thereby little influencing our results.
	In the remaining case, the thermal source overlapped with the non-thermal HXR footpoints (within our 7$''$ spatial resolution), and spectral fitting was done above 40 keV to eliminate any contamination by piled-up low-energy photons.

	Table \ref{table:2} is a list of all parameters obtained for each of our events.
	Subscripts 1 and 2 refer to the value of the leading and trailing footpoints, respectively (as determined by their heliographic longitude).
	In a few cases, the subscripts {\it strong} and {\it weak} were also used. 
	They refer to the value of the strongest and weakest footpoints, respectively (as determined by their 50-keV flux).

	Presenting all possible combination of scatter plots is prohibitive (they can be all found at a website\footnote{\url{http://sprg.ssl.berkeley.edu/~shilaire/FootPointProject/htmlsummaries/browser.html}}).
	Only the most relevant have been presented, but all have of course been examined, and an exhaustive table of computed correlation coefficients can be found in Section \ref{sect:obs:corr}.
	
 	\begin{table}
	\centering
	\caption{Measured and derived quantities.}
	\label{table:2}
	\begin{tabular}{c l}
		\hline
		Symbol	&	Name or decription	\\
		\hline
		$\gamma_1$, $\gamma_2$	& Spectral indices of both footpoints, as obtained by fitting \\
					& a power-law using the OSPEX from the {\it Solarsoft} suite of\\
					& routines \\
		$\Delta\gamma$		& Spectral index difference $\Delta\gamma={\gamma}_1-{\gamma}_2$ between \\
					& footpoints \\
		$\bar{\gamma}$		& =$\frac{1}{2} ({\gamma}_1+{\gamma}_2)$: average spectral index	\\
		F$_{50,1}$, F$_{50,2}$ 	& 50-keV photon flux in both footpoints [ph/s/cm$^2$/keV]\\
		F$_{50,\mathrm{tot}}$	& =F$_{50,1}$+F$_{50,2}$, total flux	\\
		F$_{50,\mathrm{r}}$	& =$\frac{\mathrm{F}_{50,1}}{\mathrm{F}_{50,2}}$: 50-kev flux ratio between footpoints\\
		\hline
		$dt$			& Duration or accumulation time	[s]\\
		GOES			& GOES X-ray class, or flux [W m$^{-2}$] in the low 1--8{\AA} channel\\
		\hline
		Lat$_1$, Lat$_2$	& Heliographic longitude [degrees] of both footpoints. \\
					& The footpoint with the largest longitude, (or ``leading'') is \\
					& labelled ``1'', the other one ``2''	\\
		Lon$_1$, Lon$_2$	& Heliographic latitude of both footpoints [degrees]	\\
		$s$ 			& Spherical separation between footpoints [Mm]	\\
		$\alpha$		& Angle between footpoints and solar equator \\
	\hline
	\end{tabular}
	\end{table}

\section{Observations \& Discussion}\label{sect:obs}

\subsection{Spatial information}\label{sect:obs:spatial}

			\begin{figure}
			\centering
			\includegraphics[width=10cm]{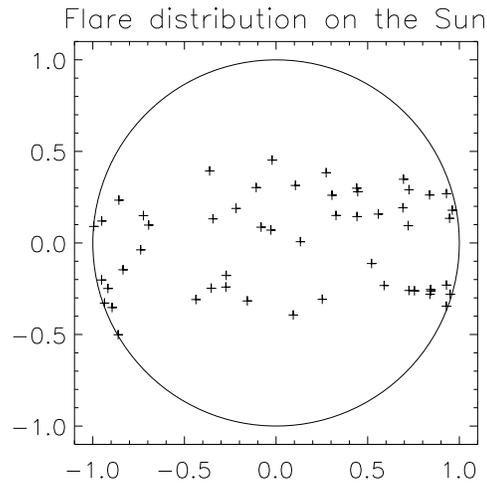}
			\caption{Distribution of our sample of 53 flares on the Sun.}
			\label{fig:FlareDistribution}
			\end{figure}

		{\bf Flare distribution on the sun:} Figure \ref{fig:FlareDistribution} shows the spatial distribution of our 53 flares on the Sun.
		As already known, flares occur predominently at $\pm$15 degrees of latitude, and there is no marked longitudinal dependence. 
		The slight lack of events at high longitudes is very probably due to observational bias: 
		with our imaging method (CLEAN with detectors 3 and above), flares with footpoint separation smaller than $\approx$10$''$ appear to be single-footpoint flare, and are not selected.
		Projection effects near the solar limb reduces the apparent footpoint separation, causing some of these flares to be discarded.
	
\subsection{Spectral Information}\label{sect:obs:spectral}

			\begin{figure}
			\centering
			\includegraphics[width=10cm]{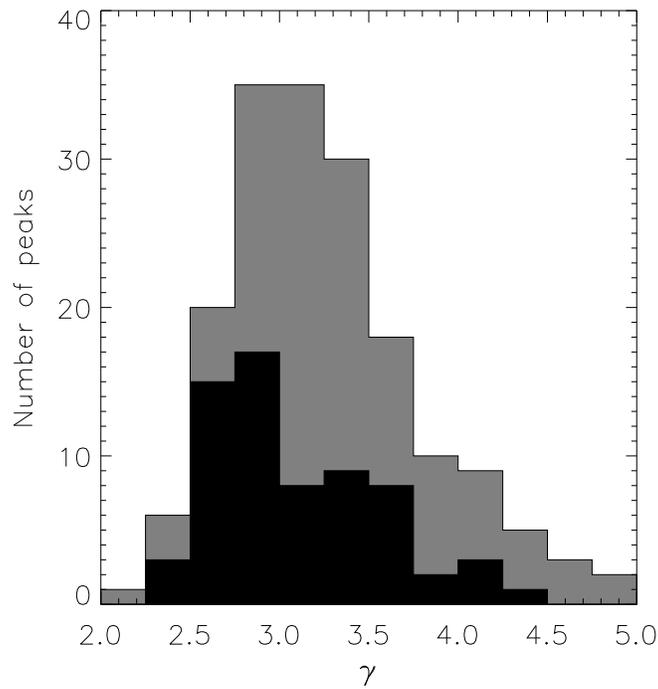}
			\caption{Histogram of spectral indices $\gamma$. The gray bars are for ``Whole Peaks'' events, the black ones for ``Peak Times'' events.
				The bin size (0.2) was taken to be larger than the average error of 0.16.}
			\label{fig:HistoGamma}
			\end{figure}
		
		Figure \ref{fig:HistoGamma} shows that spectral indices are generally harder for ``Peak Times'' events, which is of course no surprise, 
		as it is a natural consequence of the {\it Soft-Hard-Soft} behaviour observed in a majority of flares, 
		where the flattest spectral index corresponds to the time of most intense HXR emission (see {\it e.g.} \inlinecite{Paolo2004}).
		Flares are very seldom harder than $\gamma\approx$2.4 in photon spectral index (see {\it e.g.} \inlinecite{Kasparova2005} and references therein).
		The distribution at high $\gamma$ in Figure \ref{fig:HistoGamma} is not to be trusted, as it is distorted by observational bias:
		only flares with sufficient HXR emission above 50 keV , {\it i.e.} flares with hard spectra, were used in our study.

		Another observational constraint is the instrument's dynamic range $DR$:
		The weakest footpoint is visible if F$_{\mathrm{weak}} \geq \frac{\mathrm{F}_{\mathrm{strong}}}{DR}$.
		Using a conservative dynamic range of $\approx$5 for RHESSI, it means that if a footpoint is weaker than the other one by a factor 5 or more, it will not be imaged.

			\begin{figure}
			\centering
			$\begin{array}{c@{\hspace{1in}}c}
			\includegraphics[width=9.3cm]{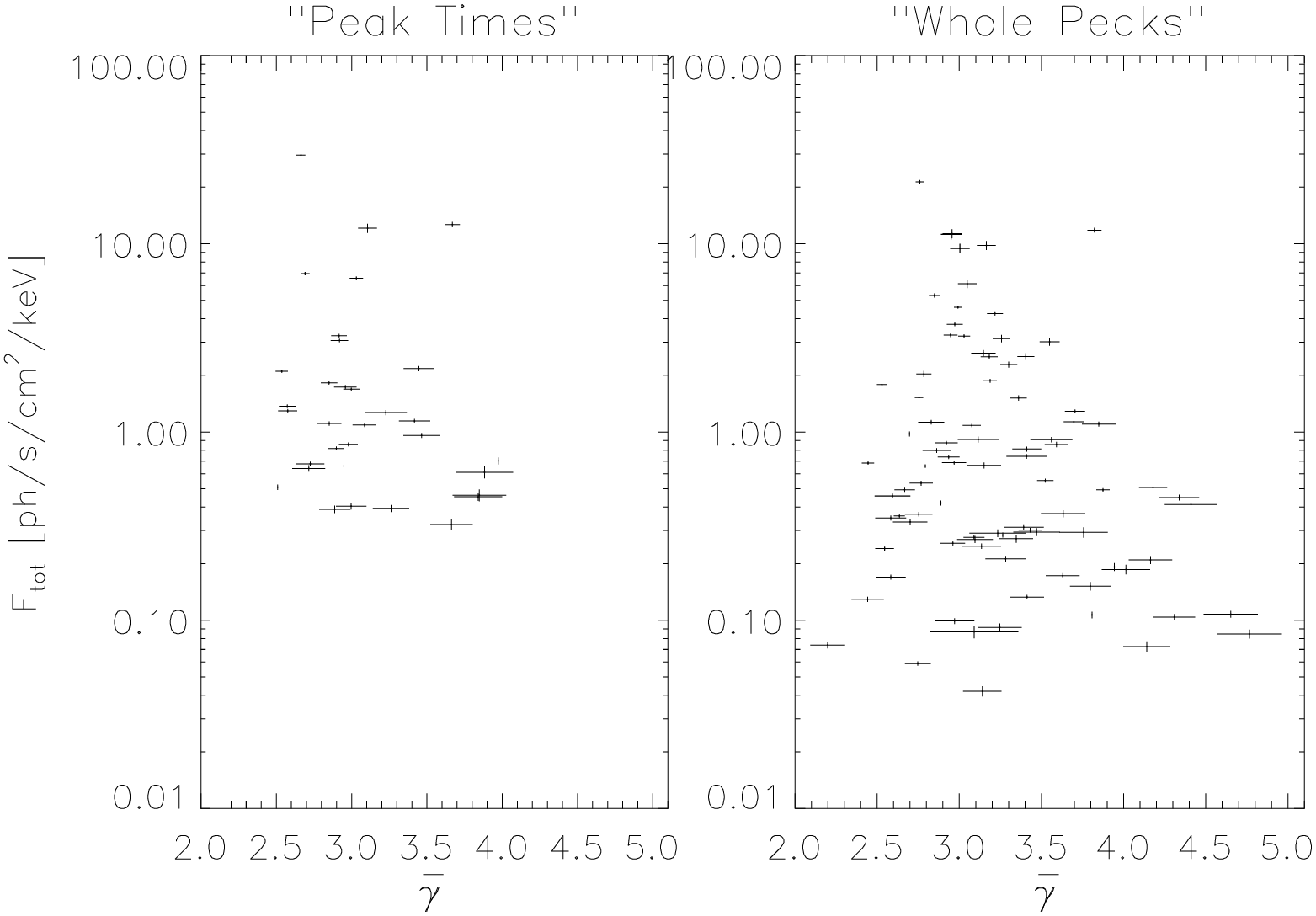}	\\
			\includegraphics[width=9.3cm]{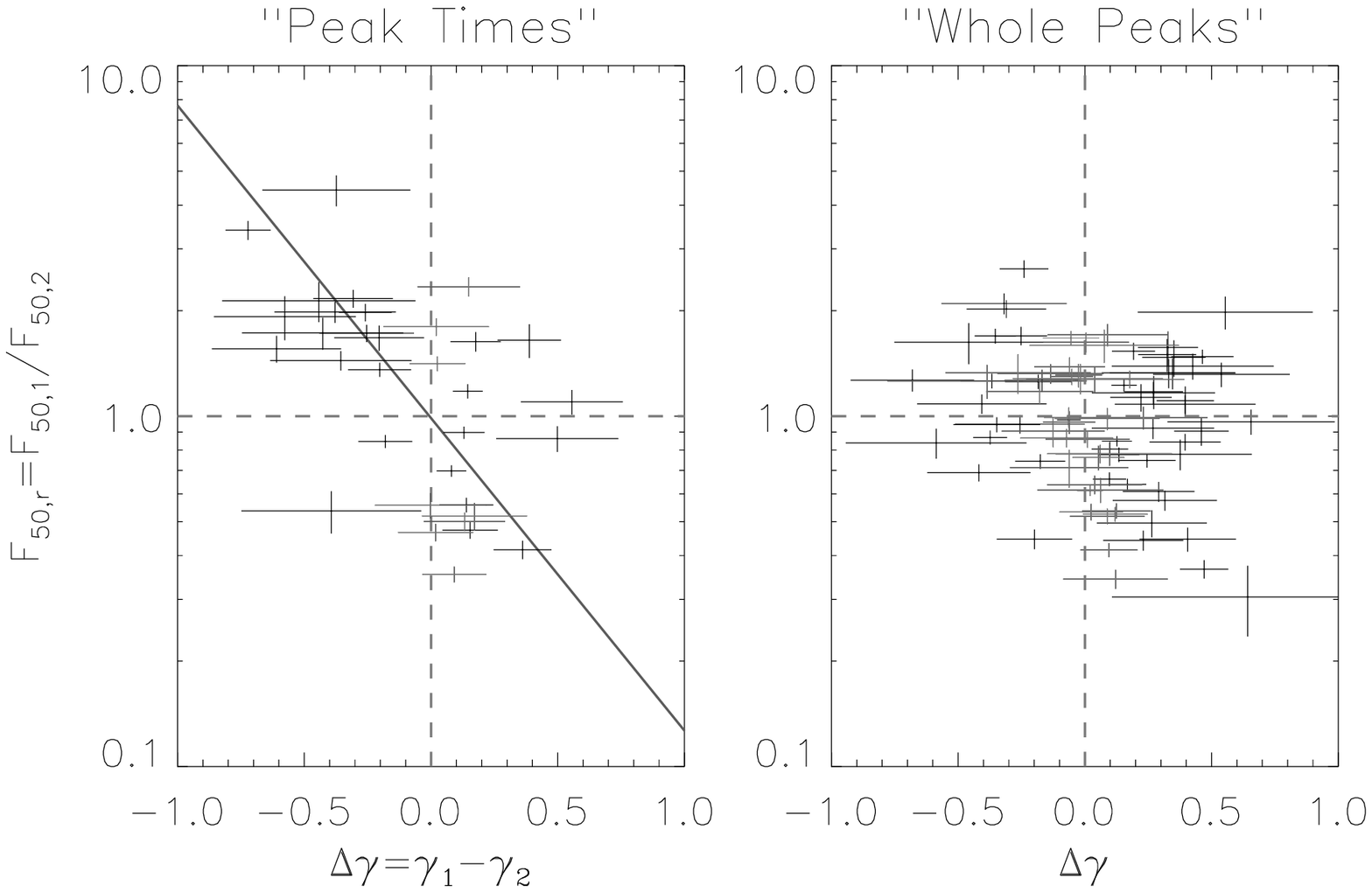}	\\ 
			\end{array}$
			\caption{Top: $\bar{\gamma}$ vs. F$_{50,tot}$, with 1-$\sigma$ error bars.
				Bottom: $\Delta\gamma$ and $|\Delta\gamma|$ vs. F$_{50,r}$, with 1-$\sigma$ error bars (see Table \ref{table:2} for an explanation of all quantities). 
				The data points in gray are the ones for which $|\Delta\gamma|<\sigma_{\Delta\gamma}$.
				The thick, gray line is a linear regression to the $\Delta\gamma$ vs. F$_{50,\mathrm{r}}$ ``Peak times'' data.}
			\label{fig:main}
			\end{figure}
			
		Figure \ref{fig:main} displays scatter plots of the average spectral indices ($\bar{\gamma}$) or spectral index differences ($\Delta\gamma$)
		versus the total 50-keV flux (F$_{50,\mathrm{tot}}$) or the  50-keV flux ratio (F$_{50,\mathrm{r}}$) of our events, with error bars.
		And indeed, no event shows a flux ratio greater than 5 (or smaller than 0.2).
		Very few footpoint pairs have spectral index difference greater than 0.6, and none above 0.8.
		This fact could not be attributed to observational effects.
		During our data reduction, a few events with $\Delta\gamma$ larger than 1 were found, but they were discarded because of poor statistics (large $\chi^2$ fitting parameter) and/or the appearance of a third source.
		
			\begin{figure}
			\centering
			\includegraphics[width=10cm]{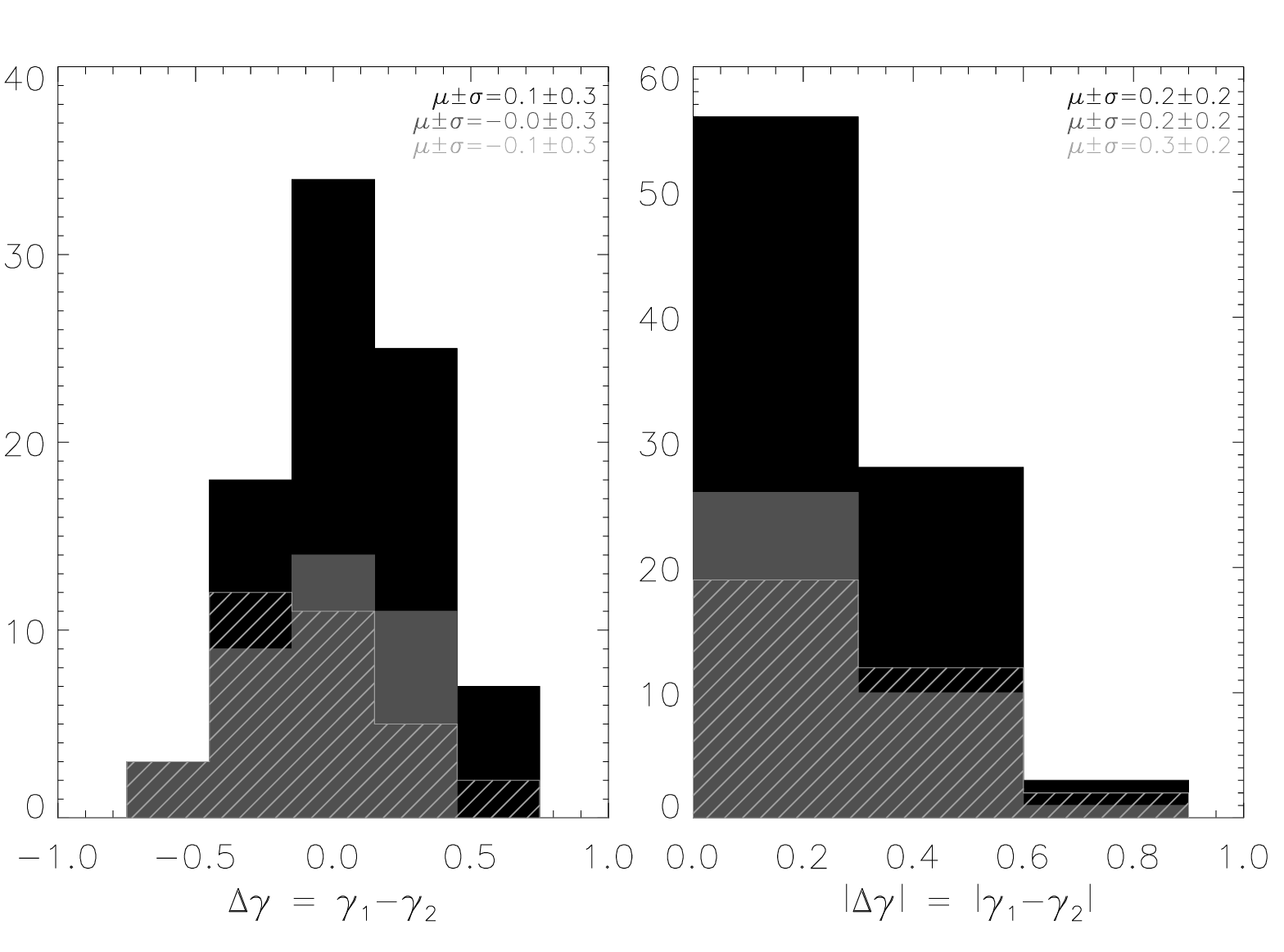}
			\caption{Histogram of spectral index differences $\Delta\gamma$.
				Crosshatched: ``Peak Time'' events. Solid black: ``Whole Peak'' events. Solid gray: ``Whole Peak'' events (only one per flare, the one overlapping the peak HXR flux time).
				The bin size (0.3) was taken larger than the average error (0.23).}
			\label{fig:HistoDg}
			\end{figure}

			\begin{figure}
			\centering
			\includegraphics[width=10cm]{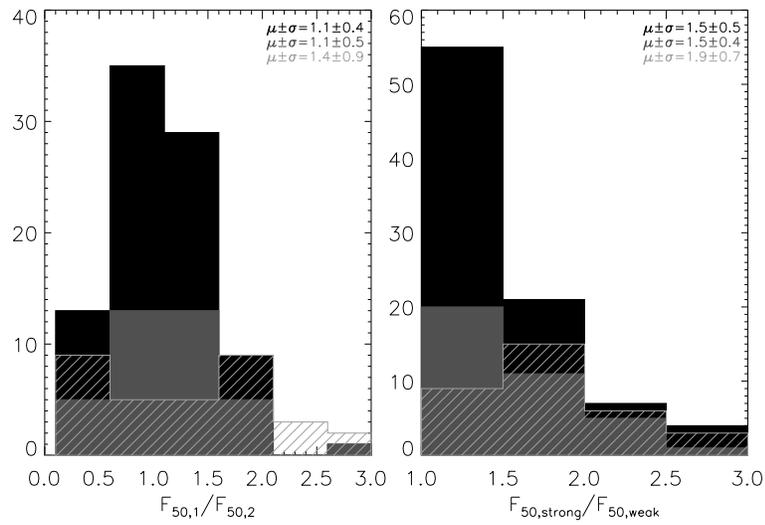}
			\caption{Histogram of flux ratios.
				Crosshatshed: ``Peak Time'' events. Solid black: ``Whole Peak'' events. Solid gray: ``Whole Peak'' events (only one per flare, the one overlapping the peak HXR flux time).
				The bin size (0.5) is much larger than the typical error.}
			\label{fig:HistoFr}
			\end{figure}

		 	\begin{table}
			\centering
			\caption{Consistency/inconsistency of $\Delta\gamma$ with zero: The numbers on the {\it left} of the ``$|$''
				are the number of cases where $\Delta\gamma - n \cdot \sigma_{\Delta\gamma}>$0, 
				and numbers on the right are events where $\Delta\gamma + n \cdot \sigma_{\Delta\gamma}<$0.
				The total number of each type of event is given in parenthesis at the top of each column.}
			\begin{tabular}{c c c c} 
				\hline
				Event type				& ``Peak Times''	& ``Whole Peaks''	& main ``Whole Peak''	\\
									& (33 events)	& (87 events)	& of each flare		\\
									& 		&		& (37 events)		\\
				\hline
				$n$=0					& 17$|$16	& 54$|$33	& 21$|$16		\\
				$n$=1, ``1-sigma results''		& 10$|$15	& 34$|$19	& 12$|$11		\\
				$n$=2, ``2-sigma results''		& 5$|$4		& 16$|$8	& 6$|$4			\\
				$n$=3, ``3-sigma results''		& 2$|$1		& 5$|$2		& 1$|$0			\\
			\hline
			\end{tabular}
			\label{table:DgPosNeg}
			\end{table}

		 	\begin{table}
			\centering
			\caption{Consistency/inconsistency of F$_\mathrm{r}$ with unity:
				Events with F$_{\mathrm{r}} - n \cdot \sigma_{\mathrm{F}_{\mathrm{r}}}>$1 are left of the ``$|$'', and  events with F$_{\mathrm{r}} + n \cdot \sigma_{\mathrm{F}_{\mathrm{r}}}<$1 are at the right. }
			\begin{tabular}{c c c c} 
				\hline
				Event type				& ``Peak Times''	& ``Whole Peaks''	& main ``Whole Peak''	\\
									& (33 events)	& (87 events)	& of each flare		\\
									&		&		& (37 events)		\\
				\hline
				$n$=0					& 20$|$13	& 41$|$46	& 21$|$16		\\
				$n$=1, ``1-sigma results''		& 20$|$13	& 40$|$39	& 21$|$15		\\
				$n$=2, ``2-sigma results''		& 19$|$12	& 32$|$34	& 16$|$13		\\
				$n$=3, ``3-sigma results''		& 19$|$11	& 24$|$32	& 12$|$12		\\
			\hline
			\end{tabular}
			\label{table:Fr}
			\end{table}

		Roughly 25\% of ``Peak time'' events (8 out of 33) have spectral index differences consistent with zero ({\it i.e.}, $\Delta\gamma$ within 1-$\sigma$ of zero; Table \ref{table:DgPosNeg}). 
		This ratio increases to $\approx$40\% for ``Whole Peak'' events (34/87).

		No ``Peak time'' event (0/33, Table \ref{table:Fr}) has a flux ratio consistent with unity, and only $\approx$10\% of the ``Whole peak'' events do (8/87; this fraction is almost reduced to zero (1/37) when considering the strongest ``Whole peak'' events of each flare ({\it i.e.} those encompassing the ``Peak time'')).
		Table \ref{table:Fr} seem to suggest that leading footpoints might have more flux during ``Peak time'' events, but the result is not statistically significant, and will not be further discussed.		

		Figure \ref{fig:HistoDg} and Table \ref{table:DgPosNeg} show that there is no statistically significant preference for the leading footpoint to be either harder or softer than the trailing one, during either ``Peak Times'' or ``Whole Peaks''.		

		The deficit of F$_{50,\mathrm{r}}\approx$1 events during ``Peak Times'' is particularly clear in Figures \ref{fig:main} (bottom left) and \ref{fig:HistoFr} (only 1 out of 33 events is within 10\% of unity flux ratio, and only 5 out of 33 within 20\%).
		This greater footpoint asymmetry during times of peak HXR fluxes than during whole HXR peaks could indicate that individual particle acceleration episode occur preferentially in one direction of the loop at any given time, but that, on the average, particles tend to be accelerated in both directions equally.

			\begin{figure}
			\centering
			\includegraphics[width=10cm]{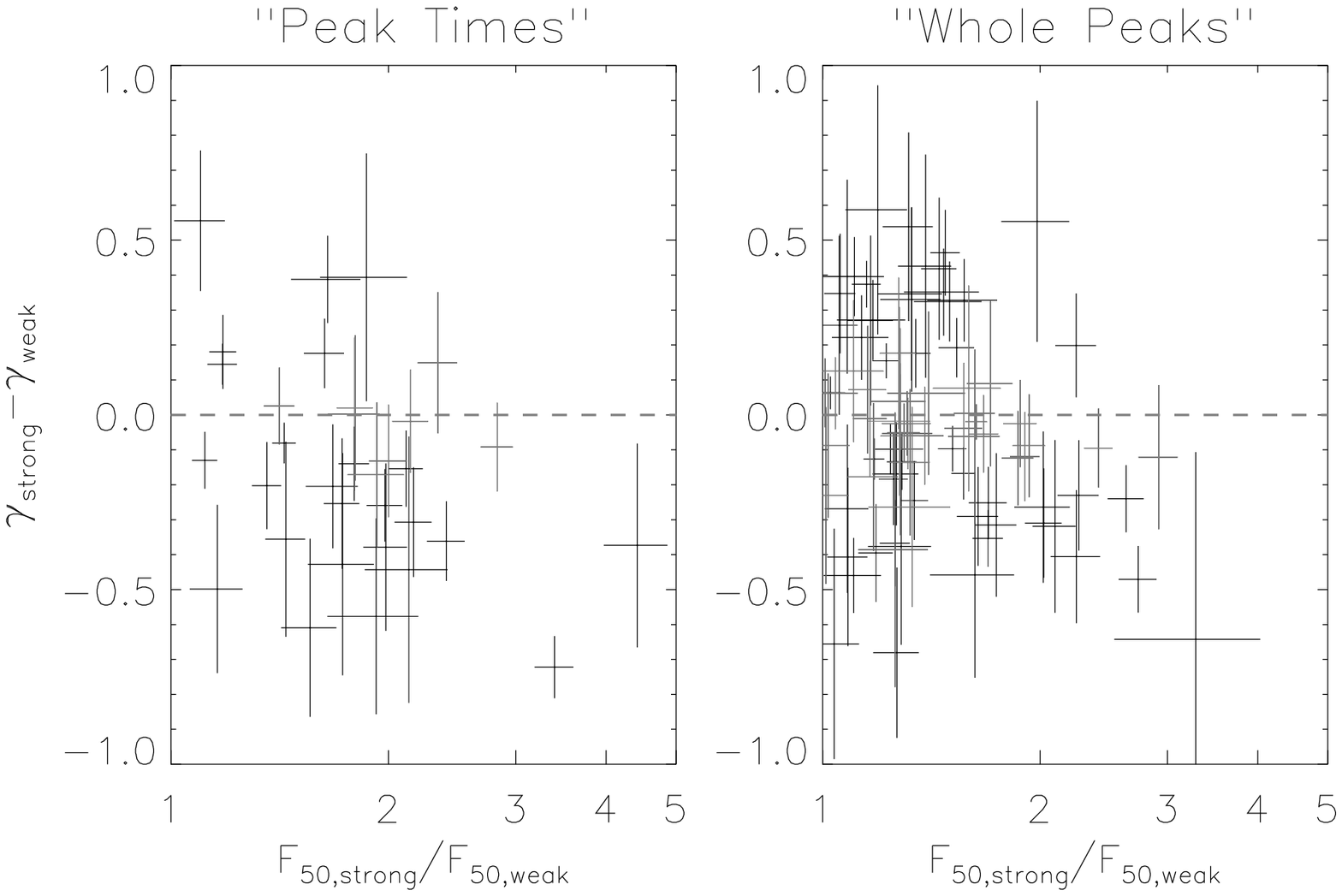}
			\caption{F$_{50,\mathrm{strong}}/F_{50,\mathrm{weak}}$ vs. $\gamma_{\mathrm{strong}}-\gamma_{\mathrm{weak}}$, where {\it strong} ({\it weak}) denotes the footpoint with the strongest (weakest) 50-keV flux, respectively.
				The dashed line marks zero spectral index difference.
				There are 23 out of 33 (79\%) ``Peak flux'' events and 53 out of 87 (61\%) ``Whole peak'' events which lie below the dashed line
				(see also Table \ref{table:ColumnDensityModelCheck}.).
			}
			\label{fig:DgVsFr2}
			\end{figure}

	 	\begin{table}
		\centering
		\caption{Column density model: (agreement/disagreement).}
		\begin{tabular}{c c c c} 
			\hline
			Event type				& ``Peak Times''	& ``Whole Peaks''	& All together	\\
			\hline
			All events				& 30.3\% (10/33)& 39.1\% (34/87)& 36.7\% (44/120)\\
			1-sigma results				& 24\% (6/25)	& 42.9\% (21/49)& 36.5\% (27/74)\\
			2-sigma results				& 28.6\% (2/7)	& 40.0\% (8/20)	& 37.0\% (10/27)\\
			3-sigma results				& 33\% (1/3)	& 60.0\% (3/5)	& 50.0\% (4/8)	\\
		\hline
		\end{tabular}
		\label{table:ColumnDensityModelCheck}
		\end{table}

	We checked in a simple way the agreement of our data set with the theory presented in Appendix \ref{appendix:ColumnDensity}: 
	that spectral index differences $\Delta\gamma$ between flare footpoints might be due to differences in column depths in asymmetric loops. 
	This effect, assuming equal distributions of electrons are accelerated in both directions of the loop, results in the footpoint having the softer spectrum also having the most flux.
	As shown in Figure \ref{fig:DgVsFr2} and Table \ref{table:ColumnDensityModelCheck}, the reverse happens most of the time, 
	{\it i.e.} better supporting a ``magnetic mirroring'' type of effect (see {\it e.g.} recent work by \opencite{Schmahl2007}, and references therein).
	Furthermore, the best candidate in support of the column density difference model is the 23 July 2002 flare (which agrees at the ``3-$\sigma$ level'') actually hits a snag when one 
	considers the amount of 50-keV flux (predicted by theory) out of the footpoints ({\it i.e.} emitted somewhere along the legs of the loop, before reaching the imaged footpoints): as explained in Appendix \ref{appendix:ColumnDensity},
	this 50-keV emission should have been observable.
	Moreover, the presence of large flux ratios F$_{\mathrm{r}}$ (2 or above, $\frac{1}{2}$ or below) also lead us to believe that this theory cannot be a dominant factor, at least for reasonable values of leg column densities (see Appendix \ref{appendix:ColumnDensity}).

	There are several altenatives to explain footpoint asymmetries:
	(a) {\it Asymmetrical acceleration:} The strong footpoint asymmetries, particularly during HXR peak times, suggest that it is the acceleration mechanism itself which could be asymmetrical.
	If the acceleration process actually took place in the chromospheric footpoints (as opposed to high in the corona), one would expect asymmetries, as both acceleration processes could in principle be independent from one another.
	(b) {\it Non-uniform target ionization:} It is conceivable that both chromospheric footpoints have different ionization structure, 
	{\it i.e.} considering the simplified step model of \inlinecite{Brown1973} or \inlinecite{Kontar2002}, that the column density required before reaching the lower-chromospheric regions of neutral atoms
	is different in both legs of the loop, perhaps due to some prior heating of only one of the footpoints. 
	Modeling and comparison with observations are required to further this idea.
	(c) The best candidate mechanism to explain footpoint emission asymmetries is magnetic mirroring, as discussed in \inlinecite{Aschwanden1999}'s 
	{\it trap+precipitation} model \cite{Melrose1976}: the magnetic field converges more rapidly in one of the footpoints, and particles are mirrored before they reach the dense lower regions (see also \opencite{Schmahl2007}).
	Our data corroborate better that scenario than the column density asymmetry model.
	The effects of {\it photospheric albedo} \cite{Bai1978} or {\it return currents} \cite{Zharkova2006} might reinforce any asymmetry observed in footpoint photon spectra, but only if the accelerated electron distributions started out different.

\subsection{Correlation Table:}\label{sect:obs:corr}

	Correlation coefficients have been computed for our data, and are displayed in Table \ref{table:corr}.

	\begin{table}
\rotatebox{90}{
\begin{minipage}[c]{15cm}
\begin{tiny}
	\centering
	\begin{tabular}{c c c c c c c c} 
		\hline
 & Duration	& 50-keV 	& 50-keV & 		  & 		   		&		 \\
 & [s] 		& flux ratio 	& flux	 & $\Delta\gamma$ & $\vert\Delta\gamma\vert$	& $\bar{\gamma}$ \\
 &&$\ln \left( \frac{F_{50,1}}{F_{50,2}} \right)$&$\ln (F_{50,1}+F_{50,2})$&&&&\\
		\hline
Duration [s] & -/1.00 & -/-0.08 & -/-0.01 & -/-0.07 & -/-0.11 & -/-0.21 \\
$\ln \left( \frac{F_{50,1}}{F_{50,2}} \right)$  &  & 1.00/1.00 & 0.06/0.12 & {\bf -0.53}/-0.23 & 0.45/0.09 & 0.32/0.22 \\
$\ln (F_{50,1}+F_{50,2})$ &  &  & 1.00/1.00 & 0.24/0.11 & -0.22/-0.13 & -0.23/-0.24 \\
$\Delta\gamma$ &  &  &  & 1.00/1.00 & -0.42/0.16 & -0.41/-0.08 \\
$\vert\Delta\gamma\vert$ &  &  &  &  & 1.00/1.00 & {\bf 0.50}/0.49 \\
$\bar{\gamma}$ &  &  &  &  &  & 1.00/1.00 \\
Longitude &  &  &  &  &  &  \\
$\vert$Longitude$\vert$ &  &  &  &  &  &  \\
Latitude &  &  &  &  &  &  \\
$\vert$Latitude$\vert$ &  &  &  &  &  &  \\
FP separation &  &  &  &  &  &  \\
FP angle &  &  &  &  &  &  \\
$\ln (\mathrm{GOES flux})$ &  &  &  &  &  &  \\
		\hline
 & 		&			  &	     &				& 	FP	& FP	& GOES	 \\
 & Longitude	& $\vert$Longitude$\vert$ & Latitude & $\vert$Latitude$\vert$   & separation	& angle & class	\\
		\hline
Duration [s] & -/0.02 & -/-0.08 & -/-0.03 & -/0.10 & -/{\bf 0.52} & -/-0.05 & -/0.22 \\
$\ln \left( \frac{F_{50,1}}{F_{50,2}} \right)$ & 0.15/0.05 & -0.24/-0.12 & -0.01/0.07 & 0.43/0.10 & -0.12/-0.07 & -0.07/0.01 & 0.16/0.12 \\
$\ln (F_{50,1}+F_{50,2})$ & 0.26/-0.06 & 0.12/0.17 & 0.20/0.01 & -0.25/-0.14 & 0.04/0.00 & 0.22/0.20 & {\bf 0.88}/{\bf 0.71} \\
$\Delta\gamma$ & -0.01/-0.08 & 0.33/0.31 & -0.20/-0.19 & -0.12/-0.18 & -0.18/-0.03 & 0.05/0.25 & 0.11/0.03 \\
$\vert\Delta\gamma\vert$ & -0.01/-0.02 & -0.25/0.12 & 0.12/-0.20 & 0.19/0.13 & 0.01/-0.18 & -0.04/0.08 & -0.09/-0.07 \\
$\bar{\gamma}$ & 0.29/0.31 & -0.24/-0.16 & 0.06/-0.03 & -0.06/0.11 & -0.01/-0.42 & -0.26/-0.02 & -0.12/-0.22 \\
Longitude & 1.00 & -0.04 & 0.17 & -0.06 & -0.02 & 0.09 & 0.12 \\
$\vert$Longitude$\vert$ &  & 1.00 & -0.27 & -0.08 & 0.13 & 0.13 & 0.07 \\
Latitude &  &  & 1.00 & -0.27 & -0.08 & 0.25 & -0.03 \\
$\vert$Latitude$\vert$ &  &  &  & 1.00 & -0.18 & -0.12 & 0.02 \\
FP separation &  &  &  &  & 1.00 & -0.15 & 0.22 \\
FP angle &  &  &  &  &  & 1.00 & 0.13 \\
$\ln$(GOES flux) &  &  &  &  &  &  & 1.00 \\                             
	\hline
	\end{tabular}
\end{tiny}
\end{minipage}
}
	\caption{Table of correlation coefficient between footpoint parameters: Where there are two numbers separated by a slash, the number to the left applies to ``Peak Times'',
	and the one to the right applies to ``Whole Peaks''. Correlation coefficients with magnitude greater than 0.5 are in bold. }

	\label{table:corr}
	\end{table}


	We found the following:
	\begin{itemize}
		\item There is some degree of anti-correlation (-0.53, with the 95\% confidence interval being [-0.74,-0.22]) between $\Delta\gamma$ and F$_{\mathrm{r}}$, for ``Peak Times'' events. 
		This is a consequence of there having more events in the {\it upper left} and {\it lower right} quadrants of the lower left panel of Figure \ref{fig:main}, and has already been discussed in Section \ref{sect:obs:spectral}.
		\item $\bar{\gamma}$ and $\vert\Delta\gamma\vert$ seem also slightly correlated: indicating that $\vert\Delta\gamma\vert$ is larger when the flare is softer.
			Upon closer examination, it appears softer spectra to be a simple case of softer spectra having larger errors.
		\item An unexpected correlation --albeit weak (the 95\% confidence interval is [0.32, 0.67])-- was found between footpoint separation and event duration, for ``Whole Peaks'' events. 

			\begin{figure}
			\centering
			\includegraphics[width=10cm]{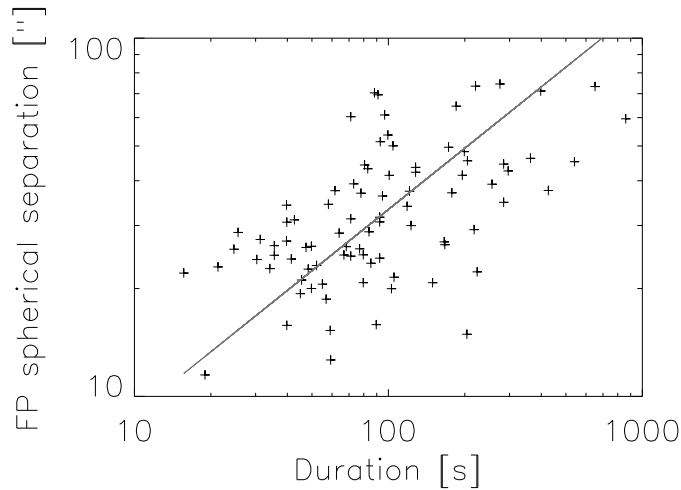}
			\caption{``Whole Peak'' duration vs. footpoint separation. The error bars are small and were not omitted.
				The gray line is a linear fit (using the bisector method) to the data.
				It yields a power-law slope of 0.6$\pm$0.1.
				}
			\label{fig:durationseparation}
			\end{figure}

		From Figure \ref{fig:durationseparation}, it seems that we have 

			\begin{equation}
				\mathrm{(HXR \,\,\, burst \,\,\, duration)} \approx \mathrm{(footpoint \,\,\, separation)^2}, 
			\end{equation}
		
		or, assuming semi-circular loops:
			\begin{equation}
				\mathrm{(HXR \,\,\, burst \,\,\, duration)} \approx \mathrm{(loop \,\,\, length)^2}
			\end{equation}
		
		It seems that the longer the loops, the longer the HXR peak will last. 
		The interpretation is not yet clear. 
		It could be a simple case of larger loops needing more time to evolve than the short ones during the flaring process,
		or another case of the ``big flare syndrome'': everything is bigger in larger flares.

		\item Excellent correlation between ``Peak Times'' total flux and GOES class (95\% confidence interval for correlation coefficient: 0.77--0.94), less good for ``Whole Peaks'' total flux and GOES class (95\% confidence interval for correlation coefficient: 0.57--0.81).
	\end{itemize}

		\begin{figure}
		\centering
		\includegraphics[width=10cm]{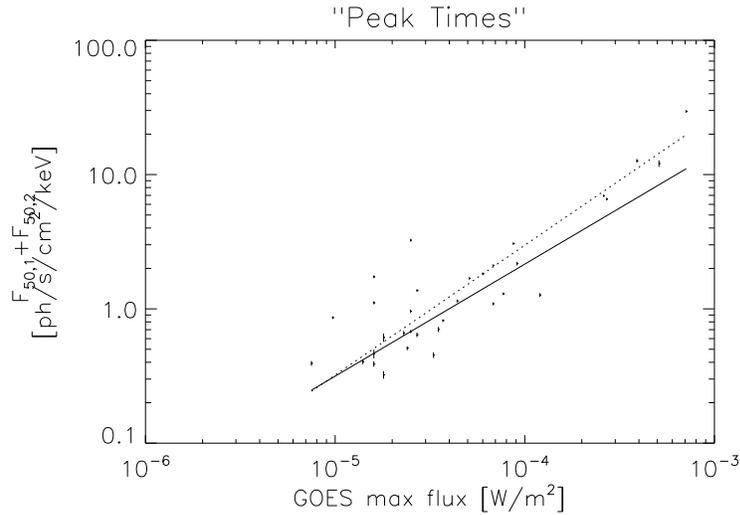}
		\caption{GOES maximum flux in the 1--8{\AA} band vs. peak 50-keV flux of both footpoints, with 1-$\sigma$ errors. 
			The solid line is a power-law linear fit to the data.
			The dotted line is another power-law fit to the data, using the bisector method.}
		\label{fig:GOESvsF50tot}
		\end{figure}

	Figure \ref{fig:GOESvsF50tot} shows the clear correlation between event's maximum GOES 1--8{\AA} flux and the 
	total HXR flux F$_{50}$, for ``Peak Time'' events. Fitting a power-law yields:
	
		\begin{equation}
			\mathrm{F}_{50} = A \cdot \mathrm{F}_{\mathrm{GOES},1-8\mathrm{\AA}}^\alpha
		\end{equation}
	where $A=(4.7\pm0.3) \times 10^3$ and $\alpha=0.8\pm0.1$ when F$_{50}$ is in photons s$^{-1}$ cm$^{-2}$ keV$^{-1}$ and F$_{\mathrm{GOES},1-8\mathrm{\AA}}$ in W m$^{-2}$.
	The {\it bisector} method \cite{Isobe1990} is more relevant when the variables are independent,
	in which case we have $A=10^{4.3\pm0.2}$ and $\alpha=0.97\pm0.05$.
	(Of course, ``Whole HXR Peaks'' events typically lie below this solid line, with a wide scatter.)

	This good correlation can be interpreted as a direct consequence of the Neupert effect \cite{Neupert1968,Dennis1985}: the larger the amount of non-thermal energy (approximated by F$_{50}$), the larger the amount of 
	thermal energy (approximated by the GOES SXR) (since F$_{50}$ is a power-law normalization factor, it describes equally well the amount of non-thermal electrons of lower energies, which contain most of the non-thermal power).
	Similar correlations have been reported before (see {\it e.g.} \opencite{Marina2005} and references therein.)
	
\section{Summary and Conclusion}\label{sect:ccl}
	The following is a compilation of our results, and can be used as a list of contraints for any flare and particle acceleration theories:

	(1) The total footpoint 50-keV flux correlates remarkably well with the GOES maximum 1--8{\AA} flux. The relationship is fairly linear.
	
	(2) There is no statistically significant difference in our sample between ``leading'' and ``trailing'' footpoints, as regards asymmetries.

	(3) Flares are mostly located at $\pm$15 degrees of solar latitude, and flare parameters have no marked longitudinal dependence.
	
	(4) Spherical separation between footpoints seem not to correlate with any of the other parameters examined, with the surprising exception of HXR burst duration, where a weak correlation was found.
	This seems to indicate that longer loops produce longer HXR peaks, probably because the magnetic disturbance and particle acceleration last longer in long loops than in short ones.

	(5) Flare footpoint spectral indices $\gamma$ are seldom below $\approx$2.4 (1 case out of 172). 
	``Peak times'' are generally harder than ``Whole Peak'' intervals, a natural consequence of the commonly observed soft-hard-soft behaviour of flares.
	
	(6) $\approx$25\% (``Peak times'') to  $\approx$40\% (``Whole Peaks'') of double footpoint flares have spectral index differences $\Delta\gamma$ consistent with zero.
	$\Delta\gamma$ can reach 0.6, and only rarely goes beyond.
	The amplitude of $\Delta\gamma$ is uncorrelated with flare GOES class.

	(7) 50-keV footpoint flux ratios are never quite unity, are typically between 1 and 2, and only seldom go beyond 3. This result could be due to observational bias.

	(8) The asymmetric loop model, where a column density difference is responsible for the difference in spectral index and flux between HXR footpoints, cannot explain a majority of our observations.
	It is therefore not a dominant factor.

	(9) The greatest asymmetry being around ``Peak Times'' further suggests that magnetic reconfiguration is greatest at those times.

\appendix   
\renewcommand{\theequation}{A\arabic{equation}}
\setcounter{equation}{0}         
\renewcommand{\thetable}{A\arabic{table}} 
\setcounter{table}{0}            

\section{Column Density Effects in Asymmetric Loops}\label{appendix:ColumnDensity}

\begin{figure}
			\centering
			\includegraphics[width=11cm]{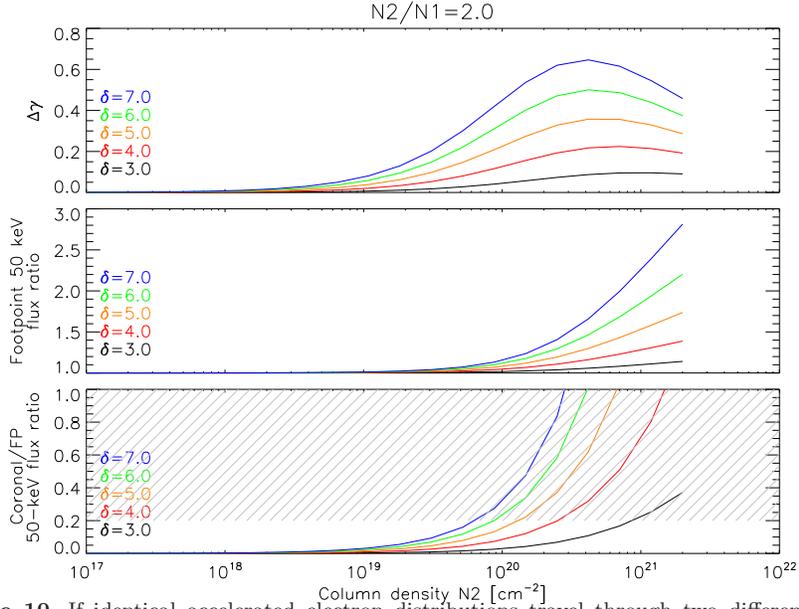}
			\caption{If identical accelerated electron distributions travel through two different column densities $N1$ and $N2$ (such as would be the case in an asymmetric loop), 
				then the bremsstrahlung photon spectra they will emit at each footpoint will be slightly different.
				Top: Spectral index difference $\Delta\gamma=\gamma_1-\gamma_2$ between footpoint spectra, assuming $N2$=2$\times N1$.
				Middle: 50-keV flux ratio F$_{50,\mathrm{r}}=\frac{\mathrm{F}_{50,1}}{\mathrm{F}_{50,2}}$ between footpoints.
				Bottom: Ratio of the total 50-keV flux present in the corona to that present in both footpoints.
				The footpoint spectral indices and 50-keV fluxes were determined by fitting a power-law in the 25--75 keV band, typical of the observations presented in this paper.
				The hatched part indicates areas where 50-keV coronal emission would be observed, assuming a conservative dynamic range of 5 for the instrument.
				$\delta$ is the accelerated electron power-law spectral index.
				}
			\label{fig:ColumnDensity1}
			\end{figure}

\begin{figure}
			\centering
			\includegraphics[width=11cm]{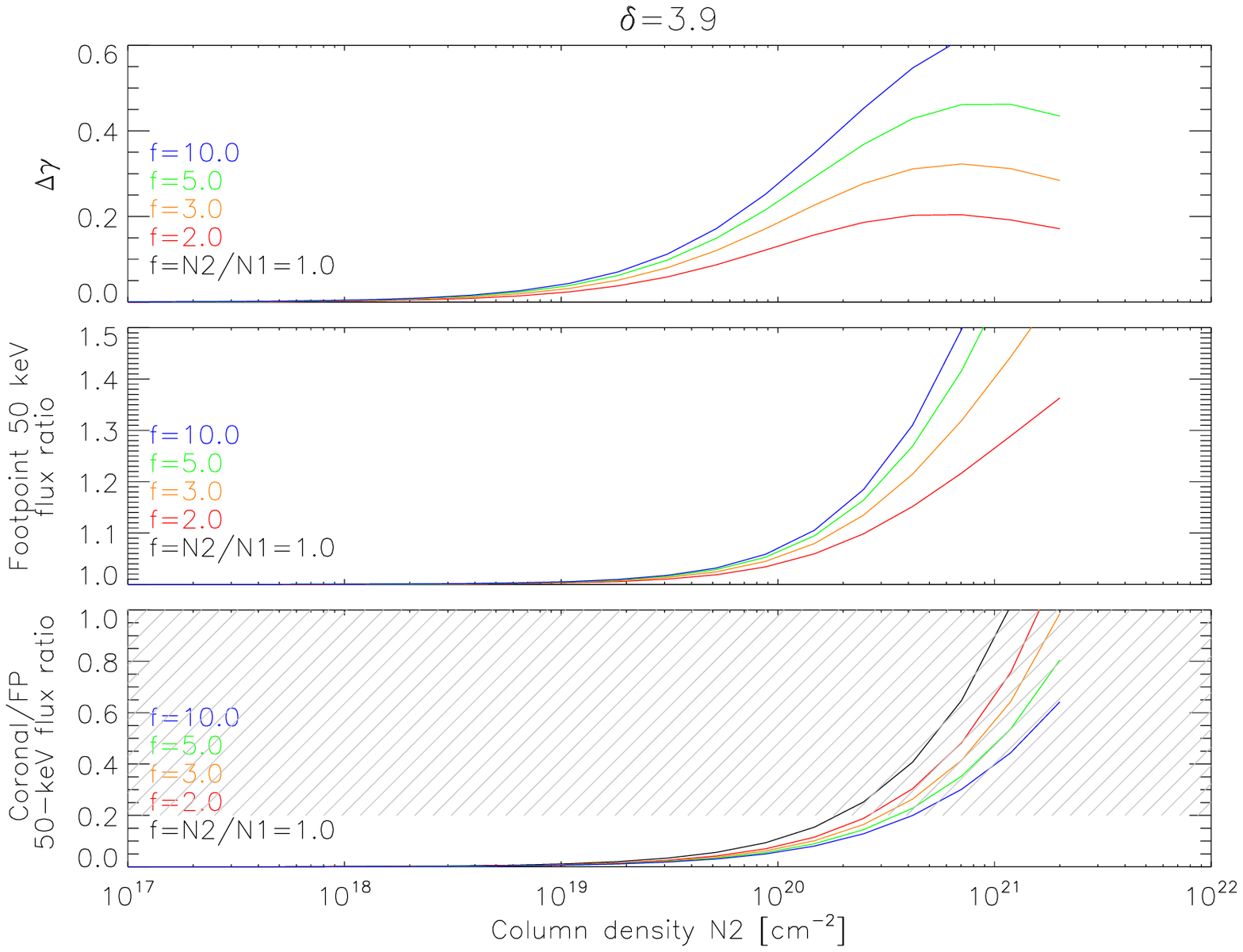}
			\caption{Similar results as for Figure ~\ref{fig:ColumnDensity1}, except that the ratio $f=N2/N1$ was varied and $\delta$ has been maintained constant at 3.9.}
			\label{fig:ColumnDensity2}
			\end{figure}

Figures \ref{fig:ColumnDensity1} and \ref{fig:ColumnDensity2} show the numerically-computed effect of loop asymmetry on the 50-keV component 
of the thick-target bremsstrahlung spectra produced by two identical accelerated electron distributions.

Notice that emission at the footpoint of the leg of the loop with the least column density is softer than the other one, but has more flux.

Increasing $\delta$ increases all three parameters ($\Delta\gamma$, F$_{50,\mathrm{r}}$, and Coronal/FP flux), whereas increasing $f$ only increases the first two, and makes the last one decrease.
Hence, in order to have large $\Delta\gamma$ and for the spatially extended coronal flux to be lost within the dynamic range of the instrument, a high $f$ (i.e loop asymmetry) is required.

%


			\begin{figure}
			\centering
			\includegraphics[width=11cm]{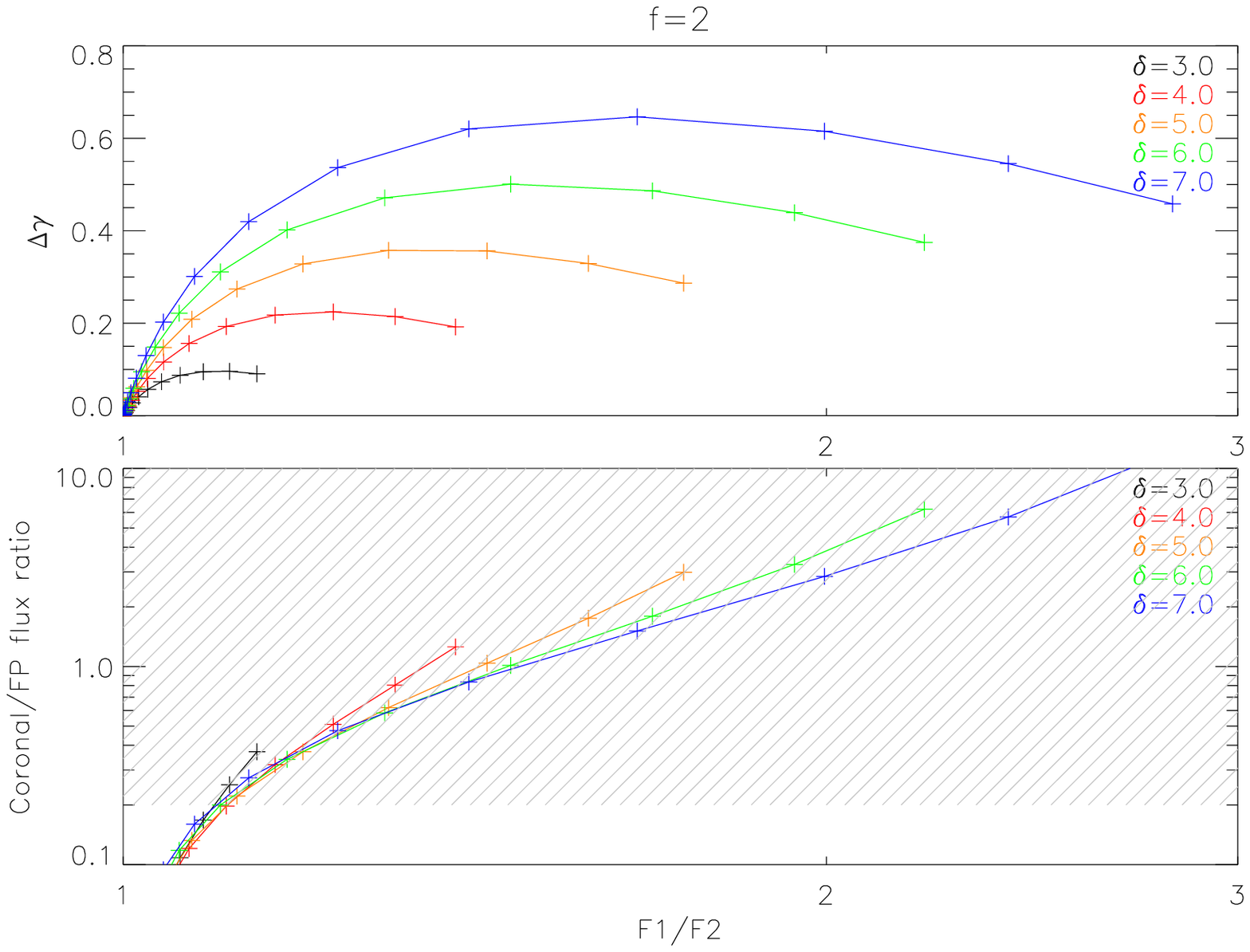}
			\includegraphics[width=11cm]{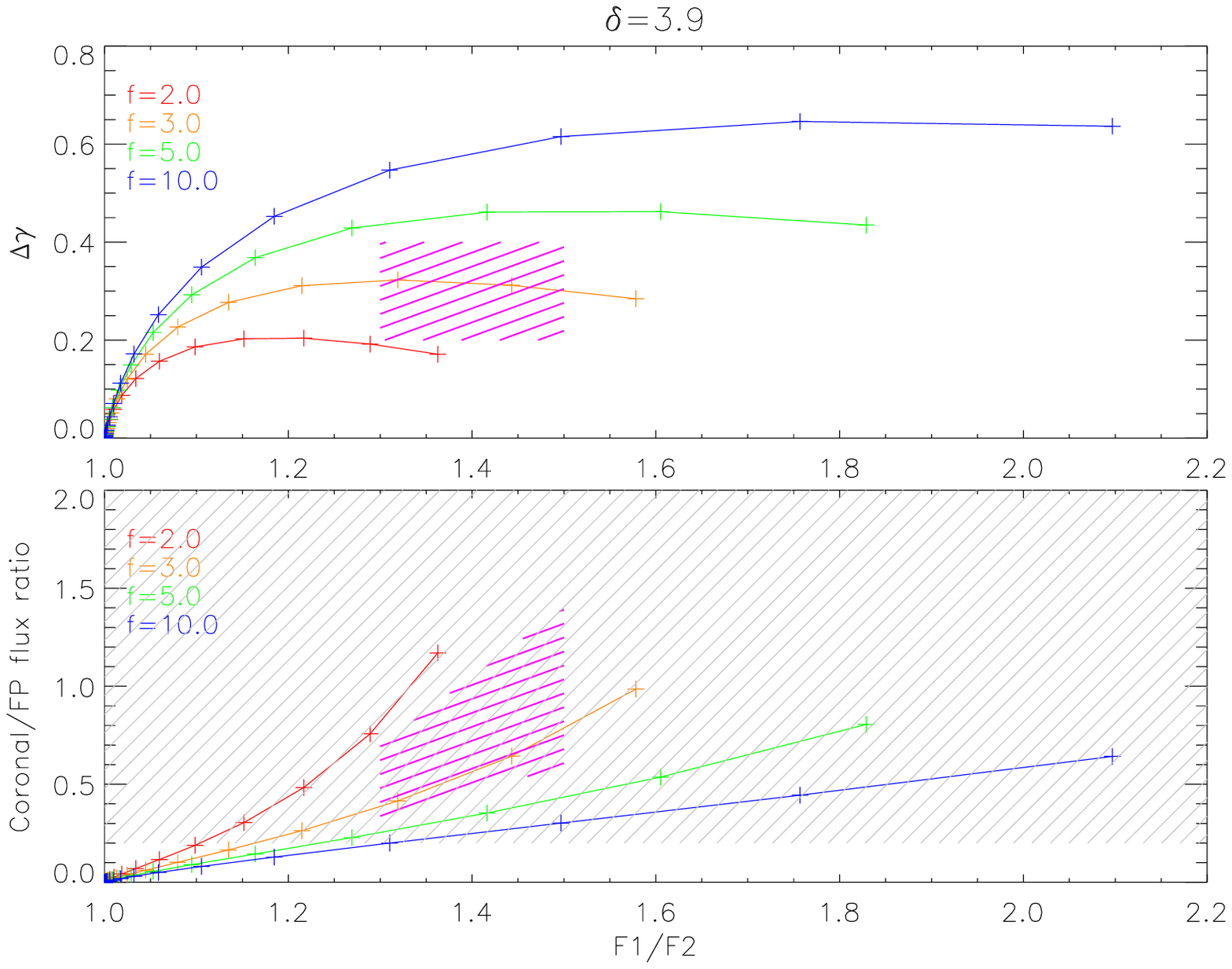}
			\caption{$\Delta\gamma$ and F$_{50,\mathrm{r}}$ for increasing $N2$. $N2$ varies from 0 (leftmost point) to 2$\times10^{21}$ cm$^{-2}$ (rightmost point).
				The purple shaded region in the third plot is the range of values for the 23 July 2002 event.
				It has been roughly mapped to the fourth plot.
				}
			\label{fig:ColumnDensity12A}
			\end{figure}

Figure \ref{fig:ColumnDensity12A} show the same data as Figures. \ref{fig:ColumnDensity1} \& \ref{fig:ColumnDensity2}, but with different axes.
The bottom plot of Figure \ref{fig:ColumnDensity12A} clearly shows that, in order for a hard flare (such as we have in this study) to have a flux ratio beyond 2 (or smaller than 1/2),
and have the coronal part of the thick-target 50-keV flux go unobserved, unreasonable values for $f$ (beyond 10!) or $N2$ would have to be considered.
In the trap+precipitation model, the footpoint with the highest (in altitude) mirror point will stop only the lowest-energy electrons.
All the hight-energy electrons will mostly stop (and emit bremsstrahlung) in the other footpoint. 
In this case, the footpoint with the hardest emission will also have the most flux.

\section{RHESSI Imaging Spectroscopy Errors}\label{appendix:errors}

	Errors in RHESSI imaging spectroscopy are extrememly difficult to estimate, as each individual pixel or feature in an image is heavily correlated
	to other parts of the image, via the point-spread function.
	
	The current heuristic method implemented in OSPEX, the standard RHESSI imaging spectroscopy software package, is to define the error
	on the flux of a feature as the maximum of the flux outside of the (visually-selected) sources, divided by a somewhat arbitrary value of $n=$3.
	When the data is very noisy ({\it e.g.} images made at high energies, were the counts are low), this method will assuredly under-estimate the errors.
	This later point is not an issue for our study, though, as we have discarted noisy images with our choice of energy bands.
	
			\begin{figure}
			\centering
			\includegraphics[width=10cm]{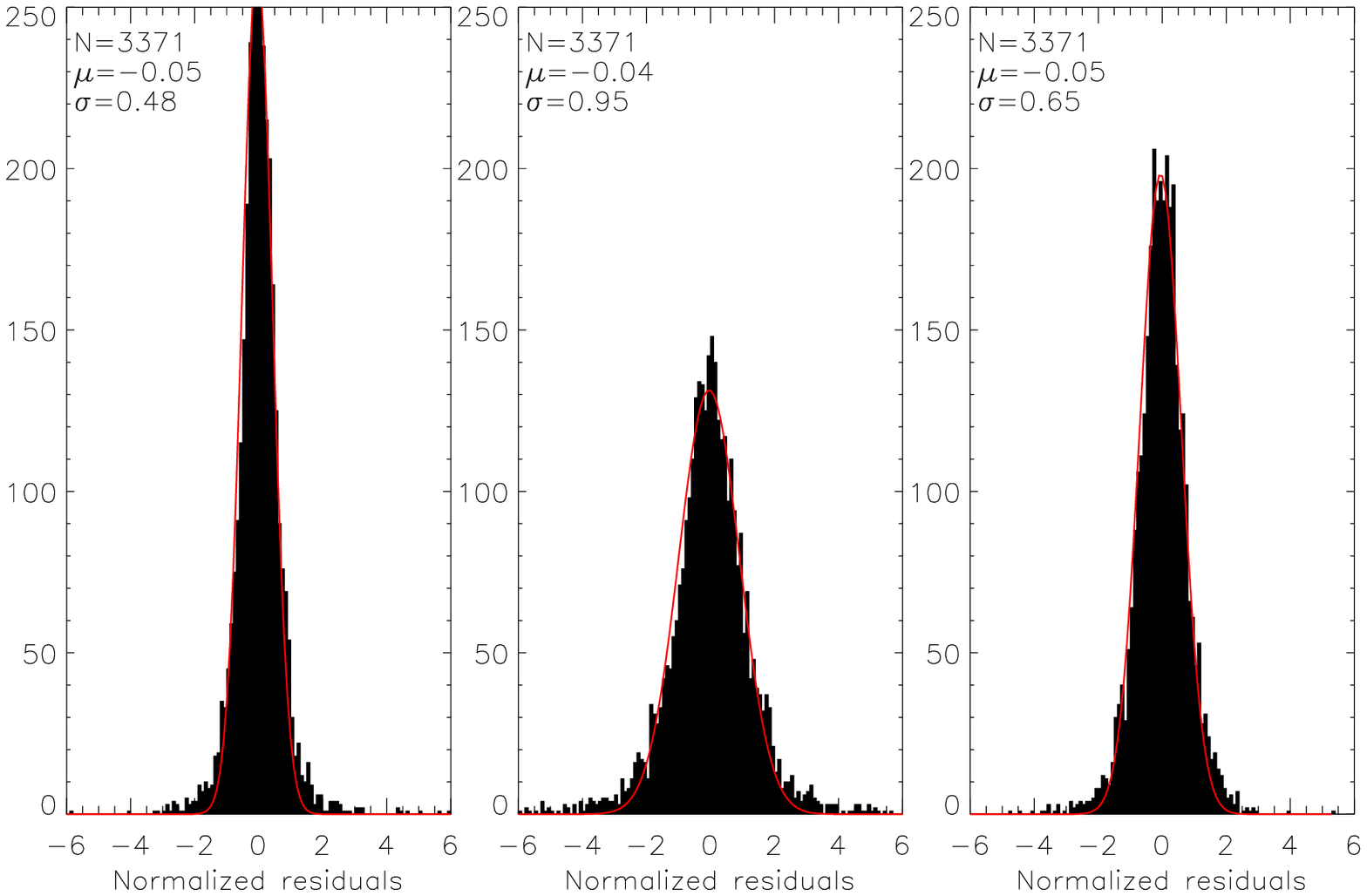}
			\caption{Histogram of normalized residuals. Left: OSPEX with $n=$3, middle: OSPEX with $n=$6, and right: Gaussian background method.
				$N$ is the total number of data points fitted, $\mu$ and $\sigma$ the mean and standard deviation of the gaussians fitted to the histograms.}
			\label{fig:normresid3}
			\end{figure}

	We have tried another method, similar to Mitani (2005), where a gaussian is fitted to a histogram of the pixel values of the whole image.
	The error on a source flux is then taken to be the 1-$\sigma$ extend of this distribution, multiplied by the number of pixels in that source.
	We have labelled that method ``Gaussian Background''.	

	To estimate the accuracy of each methods' error estimation, we have plotted (see Figure \ref{fig:normresid3}) the histogram of the normalized residuals of all our spectral fittings.
	Assuming our power-law model is correct, then the leftmost plot of Figure \ref{fig:normresid3} leads us to conclude 
	that errors in the standard OSPEX imaging spectroscopy package are typically over-estimated by a factor $\approx$2.
	We then chose $n=$6 as our heuristic number, and obtained the plot in the middle of Figure \ref{fig:normresid3}, where a fitted gaussian has $\sigma$ very close to unity.
	This is the scheme that we finally settled upon.
	
	The fitting parameters hardly change at all whether we choose to use OSPEX with $n=3$ or 6, or with the Gaussian background method, 
	only the 1-$\sigma$ error on those parameters are influenced by the choice of the method. 
	Moreover, the results and conclusions obtained using either the OSPEX with $n=$6 or the Gaussian backgound method do not change significantly.
  
\begin{acks}
	This work could not have been completed without Swiss National Science Foundation (SNSF) grant PBEZ2-108928,
	NASA Heliospheric Guest Investigator grant NNX07AH74G, and NASA contract NAS 5-98033.
	We would also like to thank the anonymous referees for their useful comments.	
\end{acks}


\bibliographystyle{spr-mp-sola}
\bibliography{../../../psh_biblio.bib}

\end{article} 
\end{document}